# Robust chirality memory in carbon nanotubes growing under modulated and evolving environments

Keigo Otsuka[1,*,†], Ryuji Fujiwara[1,†], Shigeo Maruyama[1,2,3]

[1]Department of Mechanical Engineering, The University of Tokyo, Tokyo, 113-8656, Japan

[2]School of Mechanical Engineering, Zhejiang University, Hangzhou 310027, China

[3]Institute of Materials Innovation, Nagoya University, Nagoya 464-8603, Japan

**Abstract**

Controlling the chirality and yield of carbon nanotubes is essential for their diverse applications from macroscopic composites to nanoelectronics. Floating-catalyst chemical vapor deposition is widely employed as a scalable synthesis route, where catalysts inevitably traverse spatially varying environments. However, prevailing interpretations of nanotube growth largely rely on ensemble monitoring or static, post-growth snapshots. Here, we combine digital isotope labeling with programmed temperature modulation to reconstruct the dynamic growth histories of supported individual nanotubes. Growth rates exhibit hysteresis and extraordinary sensitivity to temperature changes, indicating irreversible catalyst coarsening; nevertheless, the nanotube chirality remains robustly preserved across lengths exceeding 300 μm. This sharp contrast between kinetic adaptability and structural memory in nanotube growth indicates that chirality control can be established at the nucleation stage, while allowing independent optimization of subsequent elongation. Our findings further call for a re-examination of static interpretations of diameter-determination mechanisms derived from post-growth snapshots.

*Corresponding author. otsuka@photon.t.u-tokyo.ac.jp

†These authors contributed equally to this work

**INTRODUCTION**

Dry synthesis processes for nanomaterials can be broadly categorized into two approaches: substrate-supported processes, as represented by semiconductor epitaxy (*1*), and gas-phase processes, such as flame aerosol synthesis of nanoparticles (*2*). Viewed from the standpoint of the growing material, these processes could be regarded as Eulerian and Lagrangian descriptions, respectively. Chemical vapor deposition (CVD) growth of carbon nanotubes (CNTs) has been intensively explored through both approaches. Supported-catalyst growth benefits from strong interactions with underlying single-crystal surfaces or neighboring nanotubes, enabling highly ordered architectures (*3–5*). In contrast, gas-phase growth offers superior scalability (*6*) and compatibility with dispersion or post-separation processes (*7*), and, consequently, commercially available CNTs are often produced through continuous, floating-catalyst (FC) CVD routes, including the HiPco and eDIPS processes (*8*, *9*).

The two CNT growth modes differ in several aspects, and one fundamental difference arises from how catalyst nanoparticles grow in size: supported catalysts primarily grow through Ostwald ripening mediated by surface diffusion of atoms (*10*), whereas floating catalysts enlarge mainly through particle coalescence at relatively low temperatures (*11*). A more critical aspect specific to FCCVD processes is that each catalyst particle moves through spatially varying temperature and feedstock composition, experiencing a dynamically evolving environment rather than a fixed local condition (*12*). Whereas steady environments permit simplified descriptions of elongation kinetics (*13*, *14*), dynamic environments couple these aspects in ways that complicate rational design of growth control.

Such inherent evolution or depletion of feedstock species can alter the carbon chemical potential in gas phase and within the catalysts. Previous *ex-situ* transmission electron microscopy (TEM) and molecular dynamics (MD) simulations argue that higher chemical potential reduces the tube diameter (*15*) through a so-called perpendicular growth mode (*16*). It has also been reported that diameter changes were induced within a single CNT by temperature modulation (*17*). From a Lagrangian perspective, floating-catalyst systems inevitably experience evolving chemical potential and temperature, implying that CNT diameter should vary along the same tubes. Other MD simulations suggest that tube diameters can flexibly adjust in response to the catalyst size (*18*), further reinforcing this expectation.

However, neither raw HiPco materials nor HiPco-derived chirality-sorted samples have revealed any clear signatures of intratube junctions, despite the extensive characterizations, such as photoluminescence excitation spectroscopy (*19–21*) and transistor measurements (*22*). The preservation of chirality over hundreds of micrometers is well established for high-quality CNTs grown under steady environments (*23–25*) and supported by recent MD simulations using machine-learning force fields (*26*, *27*). This behavior has been used to infer that tube diameter should remain robust even in dynamically varying environments. Yet, has this presumed robustness ever been explicitly tested under such conditions? Recent progress has allowed us to investigate the dynamic growth history of individual CNTs under practical conditions (*28*, *29*). For instance, despite increasing the chemical potential difference $\Delta\mu$ between CNTs and catalysts more than five-fold through acetylene addition (*14*), or repeatedly reversing its sign *via* water addition (*30*, *31*), switching of diameter or growth modes was hardly observed.

A key untested question is how robustly CNTs maintain their chirality against changes in temperature and catalyst size at the single-tube level. Ensemble measurements inherently lack access to chirality continuity along each nanotube, and even analyses of isolated CNTs in *ex-situ* experiments cannot directly associate intentional perturbations with specific growth locations. Moreover, catalyst coarsening proceeds continuously and imperceptibly, with random magnitudes and directions (*10*), making it challenging to correlate catalyst-size evolution with chirality response. Although *in-situ* TEM could probe such dynamics, its intrinsic limitations, including extremely low pressure and electron-beam-induced damage, make it unsuitable for growing well-crystallized single-walled CNTs under practical conditions.

Here, we combine digital isotope labeling (*29*) with controlled heating-cooling sequences in order to examine how individual CNTs preserve chirality under dynamic environments. This method simultaneously allows us to probe the temperature dependence of growth rates at the single-tube level and to compare the kinetics of metallic and semiconducting CNTs, as well as their growth lifetimes. Most importantly, the growth rate can be used as a dynamic indicator for irreversible catalyst coarsening behavior. Even though CNTs experience simultaneous changes in temperature and catalyst size, we find that chirality is retained across >300 μm on average. While growth kinetics respond sensitively to dynamically changing environments, the initially determined structural identity of CNTs remains remarkably robust. This pronounced contrast between kinetic adaptability and structural memory highlights a key feature of one-dimensional (1D) nanotube growth. These results establish a solid baseline for discussing chirality-determination mechanisms with appropriate caution and suggest a practical strategy for achieving both structural control and scalable growth.

**RESULTS**

**Contrast between supported-catalyst and floating-catalyst growth**

To clarify the differences in the two synthesis approaches, we first compare the temperature dependences under otherwise comparable conditions. We synthesized aligned CNTs at 800 and 875°C from thermally evaporated Fe catalysts on a quartz substrate, using ethanol as the carbon source. Raman line scans were performed 10 μm away from the catalyst stripe. Diameters of individual CNTs were extracted from all observable radial breathing mode (RBM) peaks under three excitation wavelengths, as summarized in Fig. 1A and Fig. S2. The bottom panel shows a series of Raman spectra for the CNT arrays grown at 800°C. In the FCCVD process, ethanol and ferrocene were supplied as the carbon source and catalyst precursor, respectively, and carried into a 60-cm-long furnace using Ar with 3% of $H_2$ as the carrier gas. We synthesized CNTs at four different temperatures and collected them downstream on a filter, and evaluated the ensemble diameter trends from RBM spectra under three different excitation conditions (Fig. 1B and Fig. S1).

Both growth methods share the same trend, in which higher temperature shifts the diameter distribution toward larger values; however, the magnitude of this shift differs significantly (Fig. 1C). Whereas supported-catalyst growth at 900°C yielded only negligible amounts of CNTs (data not shown), FCCVD produced CNTs with an average diameter below 1.2 nm even at 950°C. Despite the use of similar Fe-based catalysts and ethanol feedstock, these contrasting temperature dependences suggest a fundamental difference in the mechanisms governing CNT diameter in the two synthesis routes.

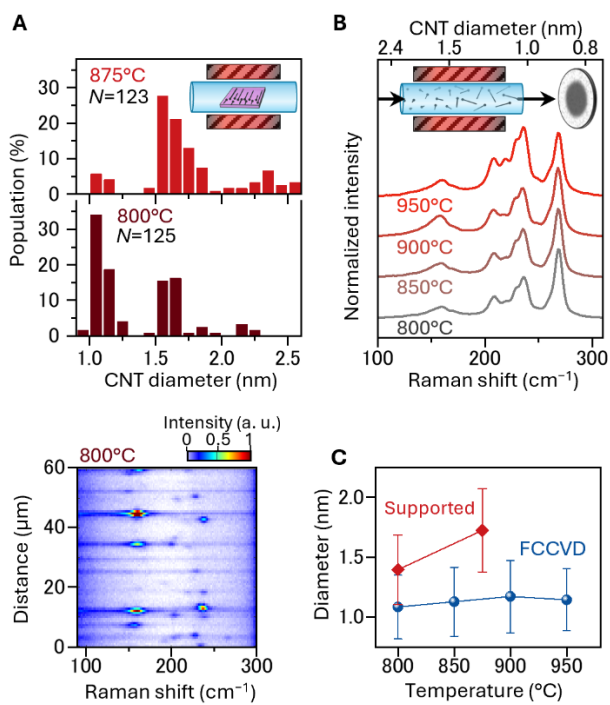

**Fig. 1. Temperature dependence of tube diameter for different growth systems.** (**A**) Diameter distributions of aligned CNTs grown on quartz substrates at 875°C and 800°C. Line scans were performed 10 μm away from the catalyst lines to derive the diameter of each individual CNT (bottom). (**B**) Normalized RBM spectra of ensemble CNTs synthesized by the FCCVD method at different temperatures (800–950°C). Excitation wavelength of 785 nm was used for panels A and B. (**C**) Comparison of temperature-dependent average diameter of CNTs synthesized by supported or FCCVD processes, obtained with excitation wavelengths of 532, 633, and 785 nm. Error bars represent the standard deviation.

**Trace of individual growth behavior under temperature modulation**

Among several possible explanations, we hypothesized that the contrasting diameter trends arise from the Lagrangian nature of FCCVD growth, in which each catalyst particle traverses a reactor with substantial gradients in temperature and precursor concentrations. To test this, we examined how individual CNTs respond to temporally varying temperatures. Because the effects of concentration variations in carbon feedstocks and etching agents on CNT growth have been examined previously (*14, 31*), our focus here is to trace CNT elongation under a programmed temperature profile. Aligned CNTs were synthesized on quartz substrates while the furnace temperature was programmed to follow a heating-cooling sequence between 800 and 873°C (Fig. 2A, top). We applied three-bit digital isotope labeling (*29*) throughout the growth (Fig. 2A, bottom) at 30 s intervals, enabling time-resolved tracing of tube elongation (Supplementary Note 1).

Figure 2B shows a Raman G-mode mapping image of the resulting CNT arrays transferred onto a Si/SiO$_2$ substrate, where brightness encodes the peak area $A_G$ and color denotes the G-mode frequency $\omega_G$. The Raman spectra collected along a representative tube (tube #1) are shown in Fig. 2C, exhibiting spatially uniform G- and D-mode features, with three distinct downward frequency shifts corresponding to isotope labels. The corresponding Raman spectra from each label, as well as from unlabeled $^{12}$C-derived portions, are shown in Fig. 2D. We can reconstruct the tube length at each label insertion time from the spacing between isotope labels (inset of Fig. 2E), revealing continuous growth over ~300–500 s and clear changes in slope reflecting the temperature change. Figure 2E displays the growth curves of multiple CNTs, where substantial diversity is observed in growth incubation time, termination, and growth rate.

Because metallic (m-) and semiconducting (s-) CNTs can be distinguished in this measurement based on G-mode features, we compared their growth-rate distributions within each temperature segment. Our earlier study at a fixed 800°C showed negligible metallicity dependence, whereas Zhu et al. reported that s-CNTs grow nearly an order of magnitude faster than their metallic counterparts at 1020°C in a kite-growth system (*32*). This motivated a systematic examination of temperature dependence using isotope labeling. In Fig. 2F, open symbols represent the growth rates of s-CNTs (blue) and m-CNTs (red) extracted for each temperature interval, while diamonds denote the corresponding mean values. Across all temperatures examined, no significant metallicity dependence is observed. These results suggest that the previously reported volcano-type bandgap dependence may emerge only at substantially higher temperatures or under specific growth modes such as kite-growth.

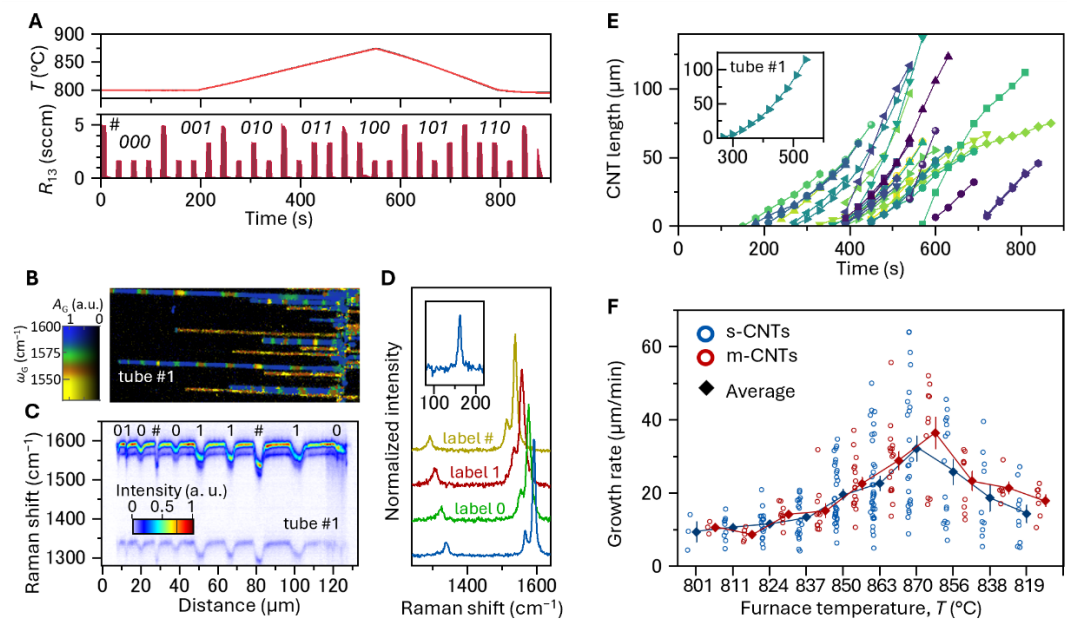

**Fig. 2. Tracking dynamic growth behavior of individual tubes on substrates.** (**A**) Time profiles of furnace temperature (top) and the flow rate of $^{13}$C-enriched ethanol pulses (bottom). The total ethanol flow ($^{12}$C + $^{13}$C) was kept constant at 5 sccm. (**B**) Raman mapping image, where brightness and color represent integrated G-mode area $A_G$ and peak frequency $\omega_G$, respectively. Catalyst particles were patterned along the right edge. (**C**) Raman spectral profile acquired along tube #1 in panel (B), sharing the same horizontal axis. (**D**) G-mode spectra of the $^{12}$C-grown segment and the three isotope-labeled segments (0, 1, #). The inset shows RBM spectra of the $^{12}$C-grown segment. (**E**) Length evolution of representative CNTs during growth, with the growth history of tube #1 shown in the inset. (**F**) Distribution of growth rates for CNTs assigned as semiconducting or metallic (open circles) and the corresponding mean values (filled diamonds), plotted as a function of furnace temperature $T$.

**Metallicity-resolved analysis on growth rate and termination**

A determining factor of the final CNT yield and length is not only the growth rate but also the growth lifetime. To quantify how growth termination depends on temperature and metallicity, we estimated the termination time of each CNT by extrapolating the growth rate at the growth front to the unlabeled terminal portion down to the catalyst stripe (Fig. 3A). Histograms of the estimated termination times (Fig. 3B) show that, for both m- and s-CNTs, termination events concentrate around ~540 s, the time at which the furnace temperature reaches its maximum. This indicates that higher temperature shortens the growth lifetime, revealing a trade-off between growth rate and growth lifetime. Although the distributions of m- and s-CNTs are not identical, the differences may arise from sampling biases related to Raman intensity and finite tube length. Accordingly, we do not observe a significant change in the semiconducting-to-metallic tube ratios within the temperature range explored (Fig. S2).

To clarify the temperature dependence, we define survival probability $p$ over unit time $\tau$ (= 1 s) as

$$p(t,T) = 1 - q(t,T)\tau, \qquad (1)$$

where $q(t,T) \equiv D(t)/N(t)\Delta t$ is the instantaneous growth-termination rate, and $D(t)$ is the number of termination events occurring within a sampling interval $\Delta t$ among those actively growing $N(t)$. Figure 3C plots $p$ as a function of the furnace temperature for ramp-up (red) and ramp-down (blue) periods, also indicating more frequent growth termination at higher temperatures. Interestingly, $p$ exhibits clear hysteresis; even at the same temperature, $p$ is systematically lower during the $T$ ramp-down stage, suggesting that growth termination depends not only on the instantaneous temperature but also on the prior thermal history of the catalyst.

The temperature-dependent growth rates $\gamma$ provide additional insights into the rate-limiting steps. Arrhenius plots constructed from growth rates ($\ln(\gamma)$ *versus* $T^{-1}$) of five representative CNTs at 30 s intervals (Fig. 3D) allow extraction of apparent activation energies $E_a$ from the slopes. We found two notable features: $E_a$ varies substantially from tube to tube, and it differs systematically between temperature ramp-up (reddish symbols) and ramp-down (bluish symbols) periods. The histograms in Fig. 3E summarize these trends for m-CNTs (top) and s-CNTs (bottom) separately, using only CNTs with at least four (three for ramp-down) segments between isotope labels to minimize fitting errors. During $T$ ramp-down, apparent $E_a$ mostly falls within 1–2 eV, consistent with prior reports (*33*, *34*). In sharp contrast, activation energies extracted during heating stage are larger with a significant scatter, which is well beyond values typically reported for catalytic CVD process and the values obtained by our isotope labeling study around 800°C (*29*).

Such anomalously large $E_a$ values during $T$ ramp-up are noteworthy. Within a simple serial-rate (series-resistance) picture, the activation energy of a rate-limiting step (the slopes of the Arrhenius plots) would be expected to decrease at higher temperatures. The opposite trend observed here could be explained by two classes of scenarios: (i) growth proceeds through parallel kinetic pathways whose relative contributions shift with temperature, e.g., a transition in the dominant carbon source from ethanol to its high-temperature pyrolysis products such as acetylene, or (ii) the growth rate becomes limited by an irreversible, non-thermal factor that evolves over time, such as progressive catalyst activation and coarsening at high temperatures. Both mechanisms would appear as a large slope of Arrhenius plots upon temperature changes.

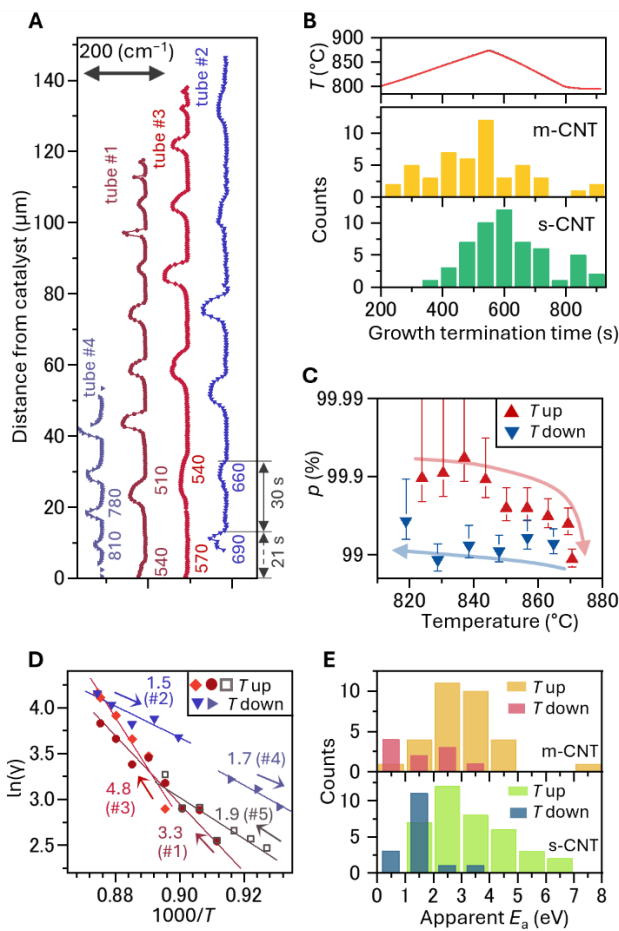

**Fig. 3. Temperature-dependent growth termination and kinetics.** (**A**) Axial profiles of the G-mode frequency $\omega_G$ for four CNTs. Numbers shown to the right of each trace indicate the onset time (in seconds) of the corresponding isotope label. (**B**) Histograms of growth termination times for m- and s-CNTs. For reference, the time evolution of the furnace temperature $T$ is shown on top. (**C**) Survival probability $p$ over a 1 s interval for a growing CNT plotted against temperature. Arrows show the direction of temperature modulation. Error bars represent the standard errors. (**D**) Arrhenius plots constructed from growth rates for five representative CNTs. Values inside and outside parentheses represent serial numbers of CNTs and activation energies $E_a$, respectively. (**E**) Distributions of apparent $E_a$ obtained by linear fitting of the Arrhenius plots, separately for temperature ramp-up ($T$ up) and ramp-down ($T$ down) periods.

**Catalyst coarsening behind the hysteresis in Arrhenius plots**

To clearly distinguish between the two scenarios above, we focused on CNTs that continued growing across both the $T$ ramp-up and ramp-down periods. This design allows the growth behavior of the same nanotube to be examined at identical temperatures but at different stages of catalyst evolution. Figure 4A compares the time evolution of CNT length for two representative cases. For the CNT that terminated growth during $T$ ramp-up period (tube #3, red symbols), the length evolution is well reproduced by a model that assumes a single $E_a$ (=4.8 eV). In contrast, for another CNT that continued growing beyond the peak temperature (tube #6, black triangles), the same single-activation-energy model fails to capture the growth behavior. Although the early-stage growth is reasonably described, the predicted length (solid line) during $T$ ramp-down significantly underestimates the experimentally observed elongation.

This breakdown cannot be explained within either a serial- or parallel-circuit picture in which the growth rate is governed solely by a temperature-dependent kinetic factor. Instead, it points to an additional, irreversible change occurring during high-temperature exposure because the growth rate was enhanced by 1.9-fold for the $T$ ramp-down period when compared at the same temperature (~837°C). As schematically illustrated in Fig. 4B, this behavior can be accounted for if the catalyst size evolves during growth, such that the growth rate scales with the catalyst surface area (*35*). Under this scenario, CNTs grown at the same nominal temperature exhibit different growth rates depending on whether they are in the ramp-up or ramp-down period, thereby reflecting the cumulative thermal history of the catalyst. Note that transition from tangential to perpendicular modes alone cannot explain this behavior because such a transition should involve the catalyst size change as long as the tube chirality remains unchanged. Kinetic bistability (*36*, *37*) can also be ruled out here because the observed $\gamma$ changes were continuous rather than abrupt, discrete switching.

To evaluate whether significant catalyst coarsening occurs under the present growth conditions, we examined the average behavior of Fe catalyst nanoparticles regardless of whether they actively nucleated CNTs. After CNT growth at 800°C and 850°C for 20 min, the nanotubes were removed by oxygen plasma, and the catalyst particles were characterized by atomic force microscopy (AFM). As shown in Fig. 4C, catalyst particles grown at 800°C exhibit little change in their size distribution over time, whereas growth at 850°C leads to the emergence of significantly enlarged particles. The size distribution of catalysts summarized in Fig. 4D exhibits a bimodal distribution (*38*) after CNT growth at 850°C as a result of Ostwald ripening, indicating nonnegligible catalyst evolution in a statistical sense.

We estimate the effect of such behavior on the growth rates of CNTs by assuming Lifshitz-Slyozov-Wagner (LSW)-type Ostwald ripening extended to sparsely distributed surface-supported nanoparticles, for which the average particle radius $\bar{R}(t)$ empirically scales as (*39*)

$$\bar{R}^4(t) - \bar{R}^4(0) = Kt. \tag{2}$$

In the system of interest, the ripening rate coefficient $K$ depends on temperature $T$ as $K(T) = K_0 \exp(-E_r/k_B T)$, where $E_r$ is the effective activation barrier for the ripening process, essentially determined by diffusion barriers, cohesive energies, and adsorption energies on the surface (*40*). The proportionality constant $K_0$ was chosen based on previous measurements that showed a ~4%/min increase of growth rates at 800°C (*41*). Note that, because $K$ is not constant over time, we use the differential form of the Ostwald ripening model ($d\bar{R}/dt = K(T)/4\bar{R}^3$), instead of Eq. 2, to obtain the time evolution of $\bar{R}$ under $T$ modulation. Using this scaling relation, we construct a deterministic model to quantify the impact of catalyst coarsening on CNT growth kinetics for the ensemble behavior. The model incorporates the programmed $T$ profile (Fig. 4E, top), $\bar{R}$ derived from the LSW-type scaling (middle), and a CNT growth rate $\gamma$ expressed as the product of a surface-area term $\bar{A} \propto \bar{R}^2$ and a thermally activated kinetic factor with an effective activation energy $E_g$ (=1.6 eV) (bottom). Thus, the overall $\gamma$ is written as

$$\gamma(T, \bar{R}) = \Gamma_0 \bar{R}(t)^2 \exp\left(-\frac{E_g}{k_B T}\right), \tag{3}$$

where the proportionality constant $\Gamma_0$ is chosen so that the obtained $\gamma$ resembles experimental values. When $\gamma$ values predicted by this model are recast into Arrhenius plots, they naturally exhibit hysteresis between the ramp-up and ramp-down branches (Fig. 4F). Importantly, this ensemble model agrees well with the average $\gamma$ for s-CNTs (Fig. 2F) when $E_r$ of 2.6 eV is used, demonstrating that catalyst coarsening can account for the large apparent activation energies and their hysteresis. Note that the data in Fig. 4F are obtained from the CNTs actively growing at each temperature, and the contributing population changed with time and significantly decreased at the late stage.

**Stochastic effects captured by kinetic Monte Carlo simulations**

The deterministic model above reasonably describes the evolution of an ensemble behavior, but does not account for the anomalously large $E_a$ for individual tubes (Fig. 4A) and the large tube-to-tube scatter (Fig. 3D). In reality, catalyst populations evolve heterogeneously; a small fraction of particles grows, while many others shrink or disappear altogether. To capture this stochastic nature, we employed kinetic Monte Carlo (KMC) simulations using the three-dimensional surface Ostwald ripening model developed by Prévot (*40*), which explicitly tracks the time evolution of individual particle radii $R$ (see Supplementary Note 1 and Figure S4 for details). Under the same temperature modulation, the size evolution of randomly selected particles is shown in Fig. 4G, and the snapshots of a single unit cell are shown on the top, where the particle diameters are enhanced for clarity. As expected, the majority of particles disappeared, while only a small fraction continued to grow, resulting in the drastic reduction in total particle number and a shift of the $R$ distribution as shown in the histogram on the right.

Because our experiments only resolve the growth history of CNTs whose length and growth rate exceed some thresholds, we assume that CNT growth preferentially occurs on relatively larger catalyst particles, and estimated CNT growth rates $\gamma$ by combining a surface area term $A \propto R^2$ with a thermally activated diffusion-precipitation term. The intrinsic activation energy $E_g$ for carbon incorporation was fixed at 1.8 eV, while growth initiation timing and lifetime were randomly determined to emulate the experimental situation. Despite this minimal parameterization, the simulated Arrhenius plots as in Fig. 4H (top) reproduce the key experimental features, namely a broad distribution of apparent $E_a$ and a systematic enhancement of apparent $E_a$ during $T$ ramp-up relative to that for ramp-down. The resulting distributions in Fig. 4H (bottom) are qualitatively consistent with those obtained experimentally in Fig. 3D, confirming that stochastic catalyst coarsening holds the key to generating both the magnitude and hysteresis of the observed $E_a$.

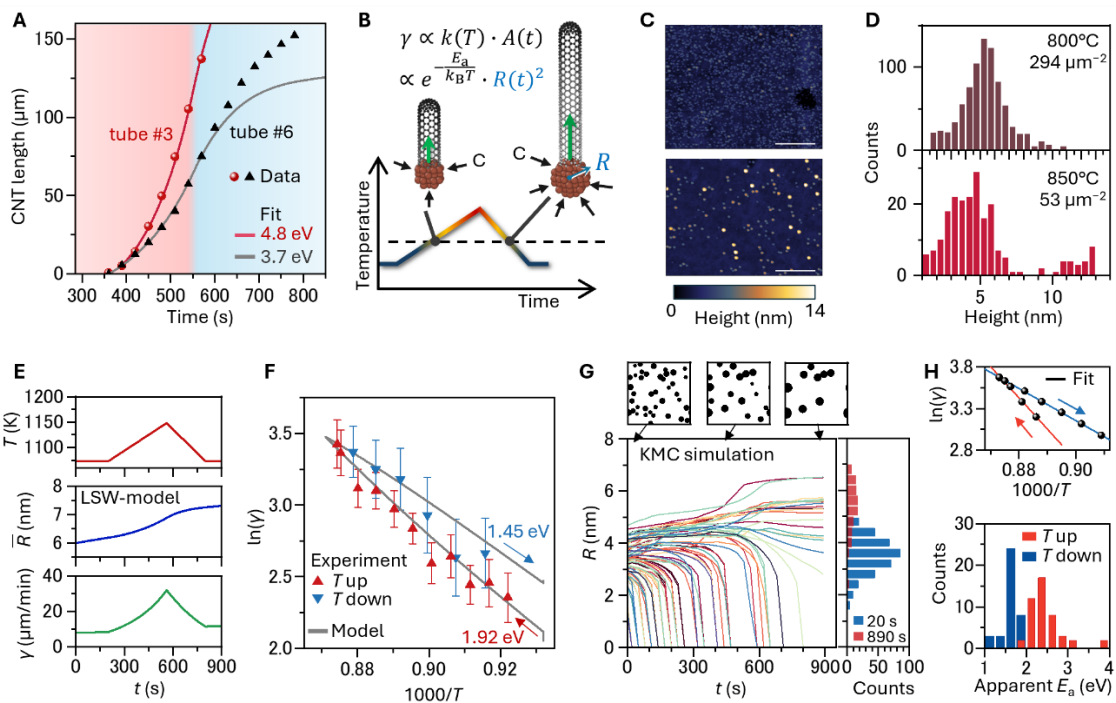

**Fig. 4. Evidence for catalyst coarsening during CNT growth.** (**A**) Time evolution of CNT length for tubes that stopped during the heating period and continued growing into the $T$ ramp-down regime. Solid curves represent predictions using a single activation energy $E_a$ fitted to the early-stage growth rates. (**B**) Schematic illustrating how discrepancies between data and single-$E_a$ fits arise when the growth rate scales with catalyst surface area. (**C**) AFM images of catalyst nanoparticles after CNT growth at 800°C (top) and 875°C (bottom) for 20 min. CNTs were removed by oxygen plasma. Scale bars: 500 nm. (**D**) Catalyst height histograms extracted from panel (C). (**E**) Programmed temperature profile (top), average radius of catalysts (middle), and corresponding growth rate $\gamma$ based on Eq. 3. (**F**) Projection of $\gamma$ to Arrhenius plots compared with average values from experiments. Error bars represent the standard errors. (**G**) Evolution of individual catalyst size obtained via kinetic Monte Carlo simulation under temperature modulation similar to panel (E). Top: snapshots for surviving particles at $t$ = 20, 530, and 890 s in a unit cell. (**H**) Recast of a representative growth profile into Arrhenius plot (top) and distributions of apparent $E_a$ (bottom), which are calculated from $T$ ramp-up (red) and ramp-down (blue) periods, when relatively large ones among simulated particles in panel (G) are assumed to grow CNTs.

A final important question is whether CNTs can preserve their diameter and chirality during elongation from catalysts experiencing Ostwald ripening. Figure 5A shows Raman profiles acquired along three representative CNTs. Along tube #3, both the RBM response and the continuity of the G- and D-mode features provide high-confidence evidence that the (*n,m*) remains unchanged along the measured length. In metallic tube #6, although RBM features are not visible, the G- and D-mode intensity, lineshape, and peak positions evolve smoothly and continuously, supporting a G,D-mode-based inference of chirality preservation. In contrast, tube #7 is a rare example in which an abrupt spectral change is observed for both G-mode and RBM. Additional full-length Raman traces (eight tubes from each category) are provided in Fig. S5. Interestingly, tube #3 fully maintains its chirality, despite a clear sign of substantial catalyst coarsening, that is a ~3.5-fold acceleration of $\gamma$ by only a ~30°C *T* increase. Likewise, tube #6 continued growing across the maximum temperature down to ~800°C without changing its chirality.

Across the full dataset, CNTs classified as chirality-preserved by RBM-based or G-mode-based criteria constitute the clear majority. Among 158 CNTs analyzed, 139 tubes (88%) maintained their (*n,m*) over the entire measured length. The total inspected length was 9.39 mm, within which 24 chirality change events were observed, corresponding to a length-normalized frequency of 2.6 mm$^{-1}$. Thus, even under pronounced environmental evolution with simultaneous temperature change and catalyst-size evolution, tube chirality and diameter are robustly preserved. Figure 5B compares this robustness with our previous experiments under steady growth at 800°C (*29*), as well as under repeated alternation between growth and $H_2O$-induced etching (*31*). Though the classification may be slightly influenced by the density of CNT arrays, the overall level of chirality preservation is reasonably comparable, indicating that dynamically evolving environments do not increase the incidence of chirality change.

Taken together, these observations support the following simple interpretation. The tube chirality is largely set at a nucleation stage and remains "remembered" during subsequent elongation under such modulated and evolving environments. Under the present 800–873–800°C protocol, most CNTs begin growth near the initial 800°C period; therefore, the diameter distribution is expected to resemble that grown at constant 800°C. To test this expectation, we investigated the diameter distribution by compiling RBM peaks from individual CNTs (Fig. 5C), in a manner similar to the analysis in Fig. 1A. As expected, the resulting distribution closely matches that obtained at 800°C under isothermal conditions (blue symbols), reinforcing the interpretation that the diameter distribution is dominated by the catalyst at the time of nucleation rather than by the later temperature history during elongation.

As a step toward extending the present interpretation to inherently dynamic growth systems, we sparsely deposited FCCVD-derived CNTs on Si/SiO$_2$ *via* the thermophoretic deposition method (*42*) while preserving as-grown structures as much as possible. As shown in Fig. 5D, Raman mapping measurements further verify the RBM continuity along nanotubes (Fig. 5E), demonstrating chirality preservation also in the floating-catalyst system.

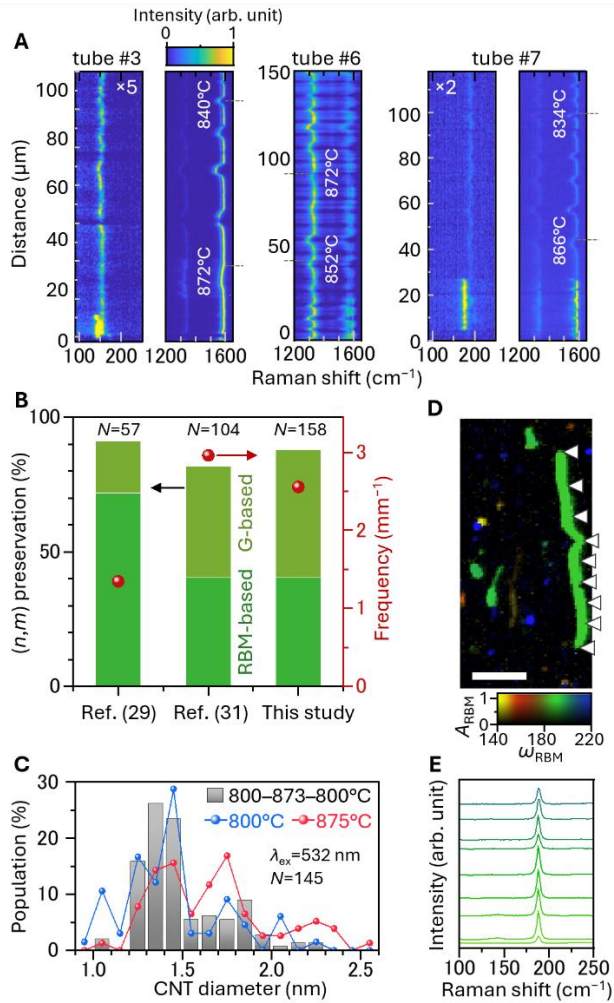

**Fig. 5. Chirality preservation under evolving environments.** (**A**) Raman profiles measured along CNTs that preserved chirality (left and center) and a CNT that exhibited local change (right). In the left panel, RBM continuity strongly supports the chirality preservation, while only G and D modes weakly support the (*n*,*m*) preservation. (**B**) Count-based fraction of chirality-preserved CNTs (bars) during temperature modulation, confirmed from RBM profiles or inferred from G-mode profiles, compared with constant-temperature growth at 800°C from Ref. (*29*) and with literature data from Ref. (*31*) involving growth interruption and $H_2O$-induced etching. Length-normalized frequency of (*n*,*m*) change is shown on the right-hand side. (**C**) Diameter distributions of CNTs measured 20 µm away from the catalyst lines for growth under temperature-modulated 800–873–800°C. Diameters were derived from RBM frequencies obtained under excitation wavelength of 532 nm. The results for constant temperatures at 800 and 875°C are shown for comparison. (**D**) RBM mapping image of FCCVD-derived CNTs directly deposited onto a Si/SiO$_2$ substrate. Scale bar: 5 µm. (**E**) RBM spectra collected along the single tube indicated by white arrows in (D).

**DISCUSSION**

We summarize key findings revealed by the supported catalyst experiments in Fig. 6A: robust retention of CNT chirality after nucleation, as well as the subsequent growth kinetics that are highly sensitive to the evolving environments. The helical and circumferentially periodic topology of the tube-catalyst interface enables screw-dislocation-like growth (*43*) without introducing crystallographic defects, which may contribute to decoupling kinetics and structural identity. This behavior stands in sharp contrast to growth in higher dimensions, where temperature and chemical potential generally influence both growth rates and crystal morphology through multiple nucleation events (*44*). With this in mind, Fig. 6B provides a conceptual framework to understand how CNT diameters are determined and can be controlled, in synthesis methods involving dynamic environments. In FCCVD, catalyst particles necessarily traverse finite axial gradients in temperature at the furnace inlet, and nucleation can therefore occur upstream once the gas-phase environment reaches a threshold temperature $T_{th}$ regardless of the set temperature $T_{set}$. Consequently, unless catalysts and carbon feedstocks are mixed only after the temperature field fully develops, the effective nucleation temperature and the corresponding catalyst size can remain relatively constant.

This framework naturally explains the weak temperature dependence of diameter distribution observed in Fig. 1. In addition, the diameter increase induced by $CO_2$ addition can be interpreted as the result of delayed nucleation under a lowered carbon chemical potential, which allows more catalyst particles to coarsen prior to nucleation, in a manner similar to $H_2$-exposure-induced diameter increase for supported catalysts (*45*) or diameter decrease induced by second carbon sources in the eDIPS process (*9*). These considerations suggest that intentional perturbations of the growth environment along the reactor axis could be exploited to confine nucleation zone, thereby narrowing the diameter distribution. We should note that the correlation between nucleation timing and CNT diameter may depend on the set temperature, because the catalyst-size evolution for $T>1000°C$ (*46–48*) can be strongly influenced by catalyst evaporation, potentially reversing the net size evolution assumed in Fig. 6A.

Beyond FCCVD-specific implications, the separation of structural determination and kinetic optimization identified here also serves as a physical basis for two-stage or seeded growth strategies, where nucleation and elongation are intentionally decoupled (*49–51*). Moreover, the relationship between CNT diameter and catalyst size observed as a snapshot after the growth should be interpreted with particular care. To establish a basis for chirality-controlled growth, tube-catalyst diameter ratios have been intensively examined using both *in-situ* (*52*, *53*) and *ex-situ* approaches (*15*, *16*, *48*). In *ex-situ* observations, in particular, one is effectively comparing a nanotube whose diameter has remained unchanged with a catalyst particle whose size has continued to evolve over time. It is therefore essential to explicitly consider how much time has elapsed since nanotube nucleation when one discusses catalyst-tube diameter relationships.

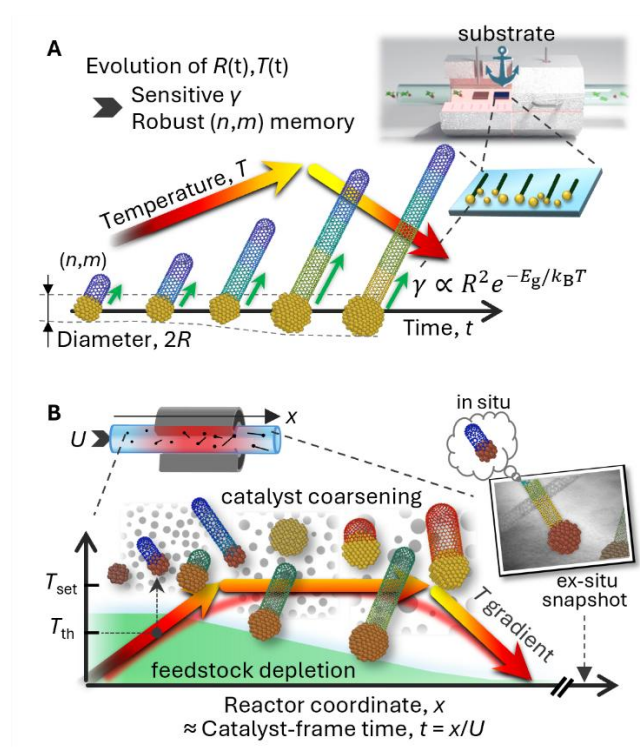

**Fig. 6. Conceptual summary of this work.** (**A**) Overview of the experimental design, key observations, and derived insights. Growth rate $\gamma$ is used to probe the unobservable dynamic behaviors of catalysts. (**B**) Schematic illustration inferred from (A) for the spatial and temporal evolution of catalysts and CNTs during FCCVD for $T<1000°C$, where catalyst evaporation can be ignored. Each CNT can initiate growth when exceeding a threshold temperature $T_{th}$ and then elongates while traversing evolving environments, including feedstock depletion, catalyst coarsening, and temperature gradients.

**CONCLUSIONS**

We have tracked the growth behavior of individual CNTs under dynamically evolving growth environments in which temperature and catalyst size vary over time. We demonstrate that, whereas the growth rate responds sensitively to environmental changes, the nanotube diameter and chirality, once determined at the early stage, are robustly preserved during subsequent elongation. The strong temperature dependence of the growth rate and the hysteresis observed in Arrhenius plots do not reflect changes in the intrinsic growth mechanism; they are attributed to irreversible catalyst Ostwald ripening that becomes pronounced at elevated temperatures. This decoupling of kinetic adaptability and structural memory is likely a characteristic feature of circumferentially closed 1D growth, suggesting that diameter and chirality control can be established at the nucleation stage, while the subsequent elongation process is independently optimized from a purely kinetic perspective. Furthermore, the present results call for a careful examination of interpretations based on *ex-situ* observations of tube-catalyst diameter ratios. A proper understanding of diameter determination requires explicit consideration of the temporal evolution from nucleation through elongation, rather than relying on static snapshots taken after growth.

**Acknowledgments**

Part of this work is supported by JSPS (KAKENHI JP22H0411, JP23K22682, JP23H05443, JP21KK0087) and JST (CREST JPMJCR20B5), the Ministry of Education, Culture, Sports, Science and Technology (MEXT), Japan. A part of this work was conducted at Takeda Sentanchi Supercleanroom, The University of Tokyo, supported by "Advanced Research Infrastructure for Materials and Nanotechnology in Japan (ARIM)" of MEXT (Proposal Number JPMXP1225UT1108).

**Author contributions**

K.O conceived the project and designed the experiments. K.O. and R.F. carried out experiments and analyzed the data. K.O. developed analytical models and performed simulations. K.O. wrote the manuscript. R.F. and S.M. commented on the manuscript.

24. R. Arenal, P. Löthman, M. Picher, T. Than, M. Paillet, V. Jourdain, Direct evidence of atomic structure conservation along ultra-long carbon nanotubes. *J. Phys. Chem. C* **116**, 14103–14107 (2012).

25. J. Gao, Y. Jiang, S. Chen, H. Yue, H. Ren, Z. Zhu, F. Wei, Molecular Evolutionary Growth of Ultralong Semiconducting Double-Walled Carbon Nanotubes. *Adv. Sci.* **10**, 2205025 (2023).

26. D. Hedman, B. McLean, C. Bichara, S. Maruyama, J. A. Larsson, F. Ding, Dynamics of growing carbon nanotube interfaces probed by machine learning-enabled molecular simulations. *Nat. Commun.* **15**, 4076 (2024).

27. S. Sun, S. Maruyama, Y. Li, Chirality-Dependent Kinetics of Single-Walled Carbon Nanotubes from Machine-Learning Force Fields. *J. Am. Chem. Soc.* **147**, 7103–7112 (2025).

28. V. Pimonov, H. Tran, L. Monniello, S. Tahir, T. Michel, R. Podor, M. Odorico, C. Bichara, V. Jourdain, Dynamic Instability of Individual Carbon Nanotube Growth Revealed by In Situ Homodyne Polarization Microscopy. *Nano Lett.* **21**, 8495–8502 (2021).

29. K. Otsuka, S. Yamamoto, T. Inoue, B. Koyano, H. Ukai, R. Yoshikawa, R. Xiang, S. Chiashi, S. Maruyama, Digital Isotope Coding to Trace the Growth Process of Individual Single-Walled Carbon Nanotubes. *ACS Nano* **12**, 3994–4001 (2018).

30. B. Koyano, T. Inoue, S. Yamamoto, K. Otsuka, R. Xiang, S. Chiashi, S. Maruyama, Regrowth and catalytic etching of individual single-walled carbon nanotubes studied by isotope labeling and growth interruption. *Carbon* **155**, 635–642 (2019).

31. K. Otsuka, S. Maruyama, Catalyst-mediated etching of carbon nanotubes exhibiting electronic-structure insensitivity and reciprocal kinetics with growth. *Carbon* **243**, 120530 (2025).

32. Z. Zhu, N. Wei, W. Cheng, B. Shen, S. Sun, J. Gao, Q. Wen, R. Zhang, J. Xu, Y.

Supplementary Materials for

**Robust chirality memory in carbon nanotubes growing under modulated and evolving environments**

Keigo Otsuka, Ryuji Fujiwara, Shigeo Maruyama

Corresponding author: Keigo Otsuka, otsuka@photon.t.u-tokyo.ac.jp

**Supplementary Note 1: Methodology details**

*Isotope labeling growth of aligned CNTs on quartz*

CNTs were grown via the alcohol CVD method using $^{13}$C-enriched ethanol feedstock as isotope labels, whose details can be found elsewhere (*1*). Briefly, CNT arrays were grown on r-cut quartz substrates using ethanol as a carbon source. Iron with nominal thickness of 0.1 nm was deposited in a lithographically defined stripes to serve as catalysts. The quartz substrate was loaded 15 cm away from the upstream furnace edge and underwent catalyst reduction treatment at 800°C in Ar/H$_2$ (3%) atmosphere at ~40 kPa for 10 min, followed by the introduction of 5-sccm ethanol to initiate CNT growth along with 50-sccm Ar/H$_2$ (3%) at the total pressure of 1480 Pa.

Ethanol with a natural isotope abundance ($^{12}$C ~99%) was used as a base carbon source, while $^{13}$C enriched ethanol (Cambridge Isotope Laboratories, Inc., 1,2-$^{13}$C$_2$, 99%, <6% H$_2$O) was introduced in a programmed manner as shown in Fig. 3A (bottom). Each $^{13}$C-containing pulse lasted for 10 s with three different $^{13}$C fractions: 33%, 67%, 100%, which are assigned as binary codes of 0 and 1, and a delimiter symbol (#), respectively. In this study, 3-bit codes such as '#000' and '#011' represent the order of introduced isotope labels, enabling unambiguous determination of absolute growth timing from three or four consecutive labels. The isotope pulses were supplied at 30 s intervals, while the furnace temperature is modulated in an ascending and descending manner (Fig. 3A, top).

*Raman spectroscopy and trace of growth history*

Raman mapping measurement was performed on the nanotubes transferred onto SiO$_2$/Si substrates. CNT arrays were transferred using poly(methyl methacrylate) (PMMA) thin film to a Si substrate with a 100 nm thick oxide layer. To locate CNTs, metallic markers (Ti and Pt) are patterned on the Si substrates prior to the CNT transfer. We use either a commercial Raman spectrometer (Renishaw, inVia) or a home-made Raman system to identify the positions and isotope fractions of isotope labels, which are then converted to length evolution of individual CNTs. The home-made system was specially designed and automatically scanned the selected regions along the CNTs whose root and tip positions are listed in tabulated data. Excitation wavelength of 532 nm was predominantly used for the sake of measurement efficiency due to the restriction of laser power and grating.

Raman spectra collected along the same tubes were fitted with Lorentzian functions, yielding the peak frequencies $\omega_G(x)$, peak heights $h_G(x)$, and peak widths $\Gamma_G(x)$ as functions of the distance from the tube root (catalyst) $x$. To determine the precise position of isotope labels, we fitted $\omega_G(x)$ at the transition from the $^{12}$C segment to the $^{13}$C-labeled segments using error functions from the large-$x$ side. This procedure allows the onset of isotope labels $x_i$, corresponding to growth times of 30 s, 60 s, etc., to be determined with minimized ambiguity. Distances between adjacent isotope labels ($\Delta x_i = x_i - x_{i+1}$) defined the instantaneous growth rates $\gamma$ at a 30-s time resolution.

*Floating-catalyst growth of CNTs*

Floating-catalyst chemical vapor deposition (FCCVD) was employed using ethanol and ferrocene as the carbon source and catalyst precursor, respectively, in a manner similar to Ref. (*2*). As shown in Figure S1A, ferrocene (0.11 wt%) and thiophene (0.01 wt%) were dissolved in ethanol, and the solution was continuously supplied at a rate of 6 μL/min using a custom-built syringe pump. The injected solution was vaporized in a heated gas line and introduced into a quartz tube reactor with inner diameter of 26 mm, together with a carrier gas flow of 500 sccm (Ar containing 3% H$_2$). The quartz tube was heated by an electric furnace with a total heated length of 600 mm. The reactor pressure was maintained at 120 kPa by pulse-width-modulation (PWM) control of a downstream solenoid valve. The as-grown CNTs were continuously collected using a downstream CNT printing system and deposited onto an area of ~1 mm in diameter on glass fiber filter paper, with a deposition time of 30 s for each condition.

*Catalyst characterization*

To investigate the catalyst size distribution, iron catalysts were deposited on r-cut quartz substrates without spatial patterning. CNTs were grown at different temperatures for the same duration and then burned via oxygen plasma at room temperature so that the catalyst size can be analyzed by atomic force microscopy (AFM). Catalyst heights were analyzed using the grain characterization tool of the Gwyddion software.

*Kinetic Monte Carlo simulations for Ostwald ripening process*

To capture the realistic impact of Ostwald ripening, a phenomenon evolving over minute-to-hour timescales, on CNT growth kinetics, we adopt a simulation framework with an appropriate level of coarse graining. Ostwald ripening of three-dimensional clusters on a two-dimensional substrate is simulated using a coarse-grained, off-lattice kinetic Monte Carlo (KMC) algorithm based on the framework of Prévot (*3*). Clusters are treated as isotropic spherical caps characterized by a constant contact angle $\theta = 90°$, and their footprint was represented by disks of radius $R$. Local equilibration of island shape is assumed to be much faster than mass transport, such that each island remains in quasi-equilibrium throughout the simulation. Periodic boundary conditions are applied in the surface plane.

The time evolution is implemented using the Bortz-Kalos-Lebowitz (BKL) algorithm, where the elementary event is the detachment of a single atom from an island. The detachment rate from island $i$ with radius $R_i$ is described by an Arrhenius-type expression including the Gibbs-Thomson correction,

$$F(R_i, L_{1,i}) = \frac{2\pi v_e a_s}{a_p \ln(L_{1,i}/R_i)} \exp\left[\frac{E_c - E_{ad} + 2\gamma_s \Omega/R_i - E_d}{k_B T}\right],$$

where $L_{1,i}$ is the distance from the island center to the nearest neighboring island edge. Other parameters are defined in Table S1. This form accounts for the probability that a detached adatom escapes the parent island and reaches a neighboring island, following the diffusion-limited off-lattice treatment.

After detachment, adatom diffusion on the substrate is treated using a first-passage coarse-graining scheme. Instead of resolving atomic-scale random walks, the adatom position is advanced by long jumps toward the nearest island, with a uniformly random angular direction. Attachment occurs when the adatom entered a capture zone of radius $d_{capt}$, at which point the atom is added to the target island. The physical time increment for each KMC step is obtained from the inverse of the total detachment rate summed over all islands.

Island coalescence is included by merging two islands whenever their separation becomes smaller than the sum of their radii plus the capture distance. The merged island is located at the center of mass, assuming strict volume conservation. Temperature is either held constant or varied in time using linear interpolation of a programmed $T(t)$ profile.

**Supplementary Note 2: Floating-catalyst chemical vapor deposition**

For FCCVD growth, the synthesis temperature was initially set to 950°C, held for 48 min at each step, and subsequently decreased in 50°C steps down to 800°C. Raman spectroscopy was performed directly on the CNTs deposited on the glass filter paper without further processing. Figures S1B and C show the RBM spectra and G-mode spectra measured using excitation wavelengths of 633 nm and 532 nm, respectively.

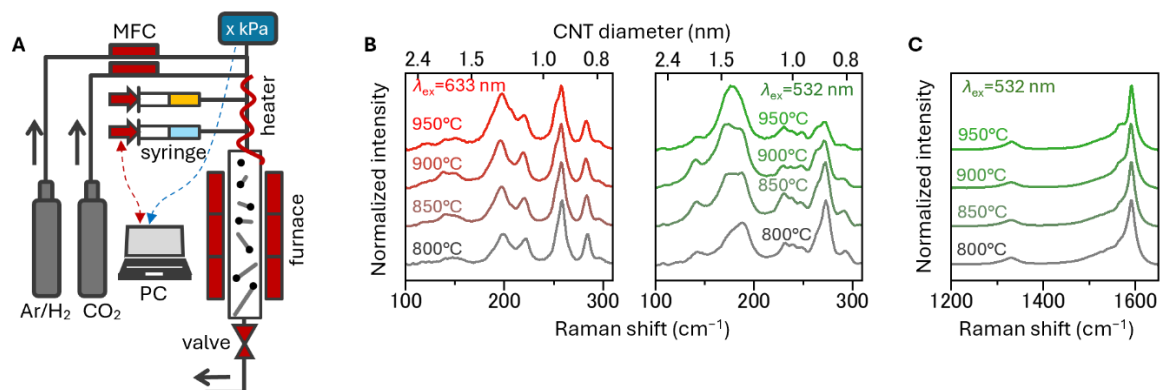

**Figure S1. FCCVD furnace and Raman analysis.** (**A**) Schematic illustration of the FCCVD reactor used in this study. (**B**) RBM spectra measured with an excitation wavelength of 633 nm (left) and 532 nm (right) for CNTs synthesized at temperatures between 800 and 950 °C. (**C**) Raman spectra of the G- and D-mode regions measured with an excitation wavelength of 532 nm.

**Supplementary Note 3: Estimation of semiconducting-to-metallic CNT ratios**

The temperature dependences of the growth rate and growth lifetime were separately examined for semiconducting (s-) and metallic (m-) CNTs, and the results are summarized in Fig. 2F and Fig. 3B, respectively. From both perspectives, s- and m-CNTs exhibited very similar temperature dependences within the temperature range examined (800–873°C). This indicates that, in this regime, there is essentially no difference between semiconducting and metallic CNTs in terms of either growth rate or growth lifetime, and therefore their relative population is expected to remain close to 2:1 at any given temperature. To directly verify this expectation, we evaluated the semiconducting-to-metallic ratio based on the detected counts of RBM peaks for samples grown at fixed temperatures of 800°C and 875°C. Using the dataset shown in Fig. 1A and spectra measured using two different excitation wavelengths (532 and 633 nm), CNTs were assigned as semiconducting or metallic within each diameter range based on their RBM frequencies (Fig. S2). For each excitation wavelength, the fraction of semiconducting CNTs is indicated in the figure. Furthermore, by weighing the number density of CNTs detected per scan length, the overall semiconducting-to-metallic ratio was estimated. This analysis yields ratios of 64:36% at 800 °C and 63:37% at 875°C, corresponding closely to the expected 2:1 ratio and showing no discernible temperature dependence. These results are fully consistent with the observation that both growth rate and growth lifetime exhibit negligible metallicity-dependent differences in this temperature range.

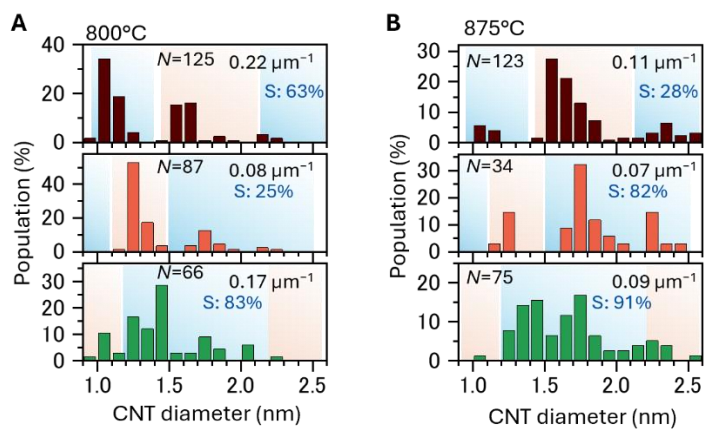

**Figure S2. Electronic-type distributions from RBM data. (A,B)** Diameter distributions of CNTs derived from Raman RBM counts for samples synthesized at 800°C (A) and 875°C (B). Metallic and semiconducting diameter regions are indicated by red and blue shaded backgrounds, respectively. The fraction of semiconducting CNTs is shown for each excitation wavelength.

**Supplementary Note 4: Temperature dependence of growth rates and lifetime**

If the anomalous acceleration of growth rate and the observed hysteresis can be attributed to catalyst coarsening at elevated temperatures, the Arrhenius plot of the growth rate is expected to deviate from linearity and become downward-convex during temperature ramp-up. This is because Ostwald ripening is accelerated at higher temperatures, which should enhance the growth rate more significantly.

Figure S3A presents Arrhenius plots of the growth rate for a larger number of individual CNTs. A weak tendency is observed in which the slope becomes slightly steeper on the high-temperature side. To further extract this feature, we focused on the heating process and plotted the apparent activation energy $E_a$ of each CNT against the average temperature of its corresponding growing duration (Fig. S3B). Although the data show considerable scatter, CNTs growing at higher temperatures tend to exhibit slightly larger apparent activation energies.

It should be noted, however, that the increase in the average catalyst particle radius $\bar{R}$ due to Ostwald ripening follows a sublinear power-law dependence on time, with an exponent of 1/4 or smaller, as discussed later. Consequently, growth acceleration in the later high-temperature stage is not expected to be abrupt but is instead governed by stochastic variations among individual CNTs. Figure S3C compares the apparent activation energies of CNTs that grew across both $T$ ramp-up and ramp-down stages, evaluated separately for each stage.

Figure S3D shows the number of growth-termination events $D(t)$ and the number of actively growing CNTs $N(t)$ as a function of time for a control experiment in which the temperature was held constant at 800°C for 24 min. Using these data, the survival probability $p$ over a unit time of $\tau = 1$ s was calculated in the same manner as in the main text, and its temporal evolution is plotted in Fig. S3E. For comparison, Fig. 3C is replotted with time as the horizontal axis. In the initial stage at 800°C, both experiments show similar values of $p \approx$ 99.9%, corresponding to a growth lifetime of ~15 min. While this high survival probability is maintained under constant-temperature conditions, a pronounced decrease in $p$ is observed when the temperature is increased, highlighting the strong impact of high temperatures on growth termination.

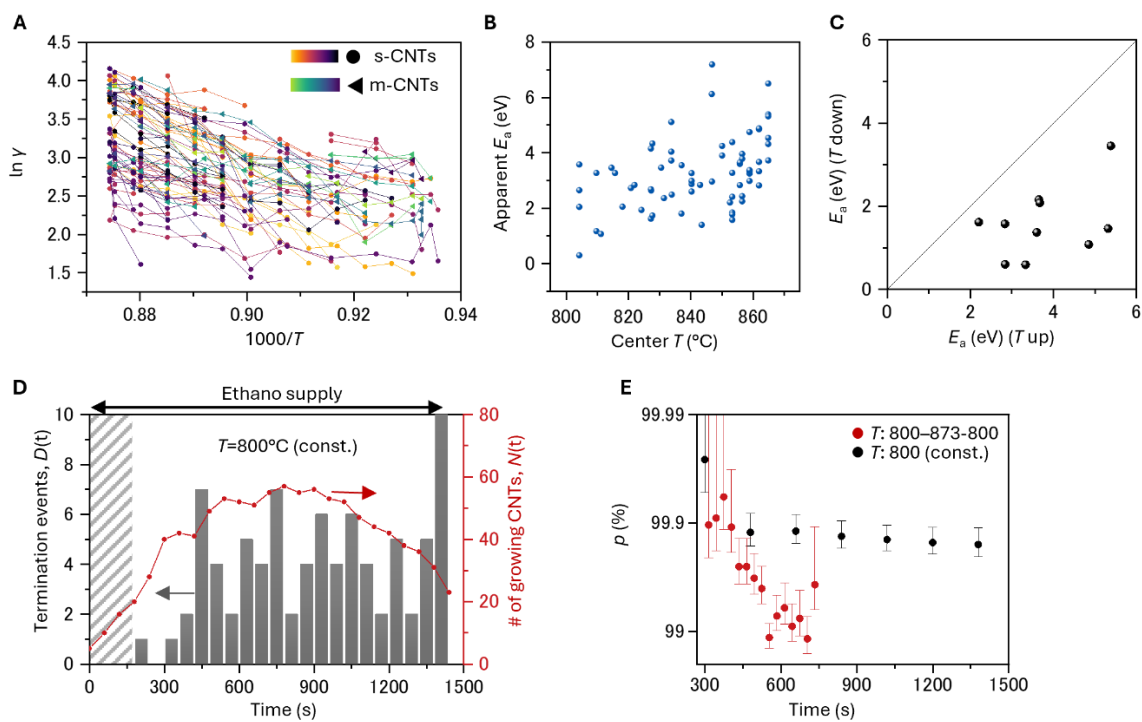

**Figure S3. Arrhenius and survival analyses supporting kinetic interpretations.** (**A**) Arrhenius plots of the growth rate for a large number of individual CNTs. (**B**) Apparent activation energy $E_a$ plotted against the average temperature of each growth segment during the temperature ascent. (**C**) Comparison of apparent activation energies $E_a$ extracted from the $T$ ramp-up and ramp-down periods for CNTs that grew across both stages. (**D**) Time evolution of the number of growth-termination events $D(t)$ and actively growing CNTs $N(t)$ for growth at a constant temperature of 800°C. $\Delta t$ is 60 s for this experiment. (**E**) Temporal evolution of the survival probability $p$ calculated from the data in (D), together with a time-replotted version of Fig. 3C for comparison. Error bars represent the standard errors.

**Supplementary Note 5: Temperature-dependent Monte Carlo simulations**

To verify the model for our kinetic Monte Carlo simulations, we run simple simulations at constant temperatures. Most of the simulation parameters are adopted from Ref. (*3*), which investigated the Ostwald ripening of Co nanoparticles on graphene surfaces, while a few parameters are adjusted to achieve better consistency with the AFM observations shown in Fig. 4C,D. All parameters used in this study are summarized in Table S1.

Each simulation initially contains 100 particles randomly distributed within a 140×140 nm unit cell under periodic boundary conditions, and each condition is simulated ten times to obtain statistical averages. Figure S4A shows time evolution of average particle radius $\bar{R}$ and the total number of all remaining nanoparticles for four different temperatures. As described by Eq. (2) in the main text, Ostwald ripening is expected to follow power-law growth behavior once the system reaches a sufficiently dilute regime. For Ostwald ripening of three-dimensional nanoparticles on surfaces, a $t^{1/4}$ growth law is widely known to hold, but it has also been reported that, in earlier stages, before particles become sufficiently sparse, the effective exponent can be smaller than 1/4. To assess the validity of our simulation model under the present conditions, where particles grow from a nominal thickness of 0.1 nm to average diameters of approximately 3–6 nm, we examined the generalized ripening relation $\bar{R}^n - \bar{R}_0^n = Kt$. Figure S4B plots $\bar{R}^n - \bar{R}_0^n$ as a function of time, where a value of $n$ = 5.5 yields an approximately linear relationship over the relevant size range. This exponent is in good agreement with previous reports for the early-stage ripening regime. Furthermore, the temperature dependence of the proportionality constant $K$ follows Arrhenius behavior, with an extracted activation energy $E_r$ of approximately 2.6 eV. We note that the value of $E_r$ = 2.6 eV was first determined from the temperature dependence and hysteresis of the experimentally measured CNT growth rates based on comparison with the deterministic ensemble model for Ostwald ripening (Fig. 4F). Since Ref. (3) considers a different system (Co on graphite) from our experiments, we then slightly adjusted a few parameters in Table S1, so that the independently performed KMC simulations yield a consistent $E_r$ through the analysis shown in Fig. S4B.

Using the same KMC model, we then simulated the size evolution of individual nanoparticles under time-varying temperature conditions, analogous to those employed in the experiments. Ensemble behaviors, such as $\bar{R}$, $\bar{R}^n - \bar{R}_0^n$, and the total number of nanoparticles are summarized in Fig. S4C,D. In addition, the trajectories for individual nanoparticles are shown in Fig. S4E. In the simulation for length evolution of CNTs, we assumed that the 50 largest nanoparticles at the final time were involved in CNT growth, while the remaining particles were neglected, as they mostly vanish during the ripening process. CNT growth was then simulated using a simplified scheme in which growth initiates and terminates at random times for each selected particle. Note that this latter assumption is not strictly accurate, because, as discussed in Fig. 3C, the probability of growth termination depends on temperature and its prior history. The nanoparticle size evolution is shown by black lines in Fig. S4E, while the time intervals assigned as active CNT growth are indicated by colored lines and symbols. The corresponding temporal evolution of CNT lengths obtained from this model is shown in Fig. S4F.

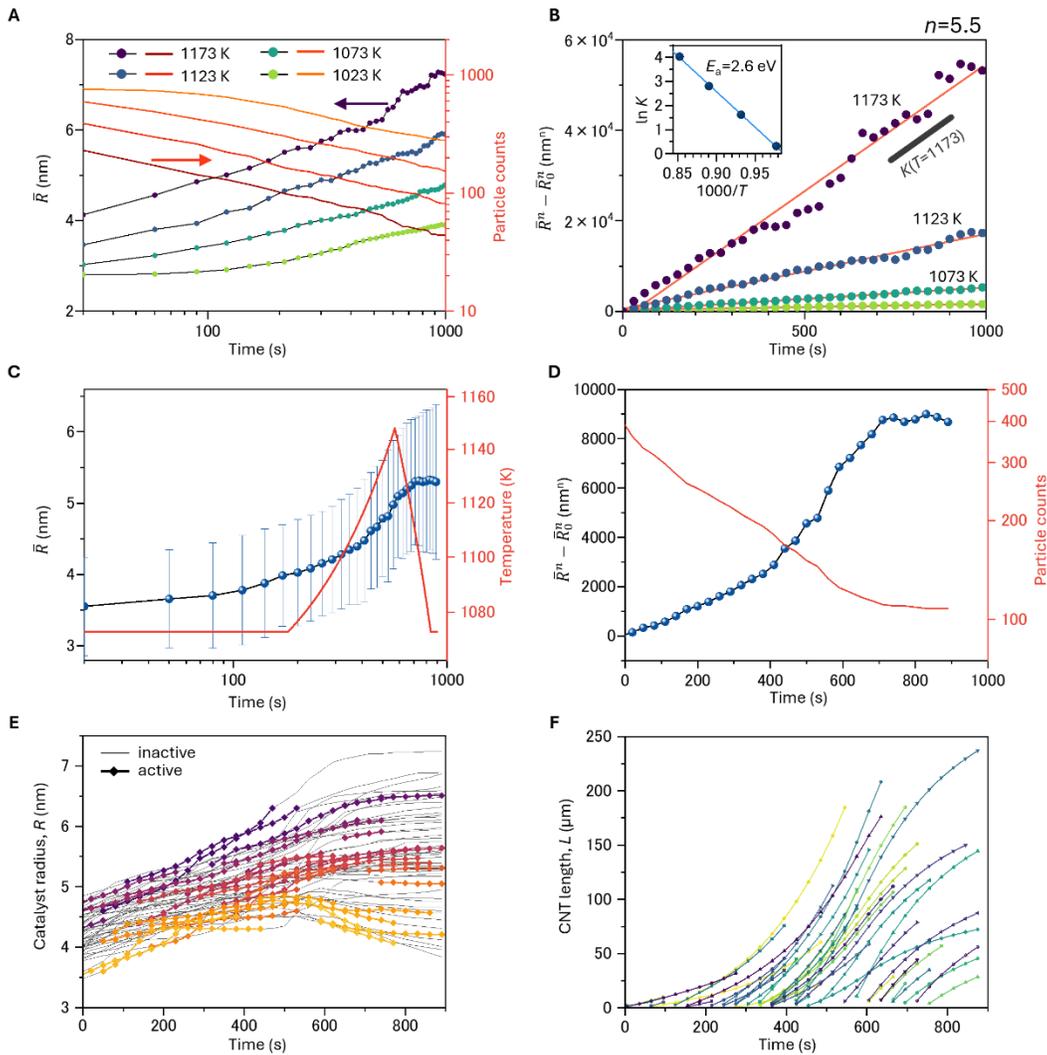

**Figure S4. KMC validation and connection to simulating CNT length evolution.** (**A**) Time evolution of average radius $\bar{R}$ and the total number of nanoparticles obtained from the KMC simulation at different temperatures. (**B**) Time evolution of $\bar{R}^n - \bar{R}_0^n$, where $n = 5.5$ provides an approximately linear relation over a given size range. Inset: Arrhenius plot of the prefactor $K$ in the Ostwald ripening model, yielding an activation energy of 2.6 eV. (**C**) Time evolution of $\bar{R}$ during temperature modulation (right-hand axis). (**D**) Time evolution of $\bar{R}^n - \bar{R}_0^n$ and the total number of nanoparticles. (**E,F**) Time evolution of individual nanoparticles that we assume to grow CNTs in Fig. 4G-H (black lines) (C) and of lengths of individual

CNTs. The periods assigned as active CNT growth in (C) are indicated with colored lines and symbols.

Table S1. Parameters used in the KMC simulations and their physical meaning

| Symbol / Parameter | Value | Unit | Physical meaning |
|---|---|---|---|
| $L_x$, $L_y$ | 140×140 | nm$^2$ | Simulation cell size in x and y (periodic) |
| $n_{int}$ | 100 | – | Initial number of islands |
| nominal thickness | 0.1 | nm | Equivalent mean film thickness |
| $\theta$ | $\pi/2$ | rad | Contact angle of spherical-cap islands |
| $\Omega$ | $0.25^3$ | nm$^3$ | Atomic volume |
| $\gamma_s$ | 2.5 | J m$^{-2}$ | Surface free energy of the island |
| $\nu_e$ | 1×10$^{-14}$ | s$^{-1}$ | Attempt frequency for atom detachment |
| $a_s$ | 0.246 | nm | Substrate lattice spacing |
| $a_p$ | 0.25 | nm | Atomic spacing at island perimeter |
| $E_c$ | −4.386 | eV | Cohesive energy of island material |
| $E_{ad}$ | −1.6 | eV | Adsorption energy of adatom on substrate |
| $E_d$ | 0.1 | eV | Diffusion barrier on substrate |
| $d_{capt}$ | 0.246 | nm | Capture radius for adatom attachment |

*All parameters are chosen consistently with the off-lattice KMC formulation for diffusion-limited Ostwald ripening described in Ref. (3) with slight modifications for $\nu_e$ and $E_{ad}$.*

**Supplementary Note 6: 1D Raman spectra along CNTs**

According to the definition by Gao et al. (*4*), where a generation is defined as the height of a single atomic layer growth (~0.12–0.14 nm), it is estimated that the CNTs maintain their chirality for an average of three million generations, even under fluctuating environmental conditions such as variations in catalyst size and temperature. This chirality preservation along individual nanotubes was investigated by Raman spectroscopy measured along the same nanotubes in 600 nm intervals using an excitation wavelength of 532 nm, covering >15,000 points in total (9.39 mm). Figure S5 presents representative Raman maps for 24 CNTs, which are classified into three categories based on the confidence level of chirality assignment. The first group consists of CNTs that show clear chirality preservation over their entire lengths, as evidenced by the continuity of RBM features (Fig. S5A). The second group exhibits chirality preservation with lower confidence, where continuity is inferred mainly from the G- and D-mode features in the absence of clearly resolved RBMs (Fig. S5B). The third group includes CNTs that display apparent intratube chirality changes, manifested as discontinuities in RBM frequencies or concomitant changes in the G- and D-mode spectra (Fig. S5C).

For the CNT that exhibits a chirality change (tube #7), we examined the timing of the transition, the associated diameter change, and its impact on the growth rate. The chirality change occurs near the maximum temperature and is most plausibly assigned to a transition from (15,2) to (15,5). Immediately after the chirality change, the growth rate decreases markedly; nevertheless, within each segment before and after the transition, the temperature dependence of the growth rate remains consistent with the trends discussed throughout the main text. Figure S6C summarizes the chirality-change timing for ten CNTs together with the apparent activation energy extracted from the growth segment preceding the transition, which is related to the degree of catalyst coarsening. Although the number of samples is limited, chirality-change events appear to occur more frequently near the maximum temperature of the modulation profile. No clear correlation between the apparent activation energy and the transition timing is observed within the present dataset.

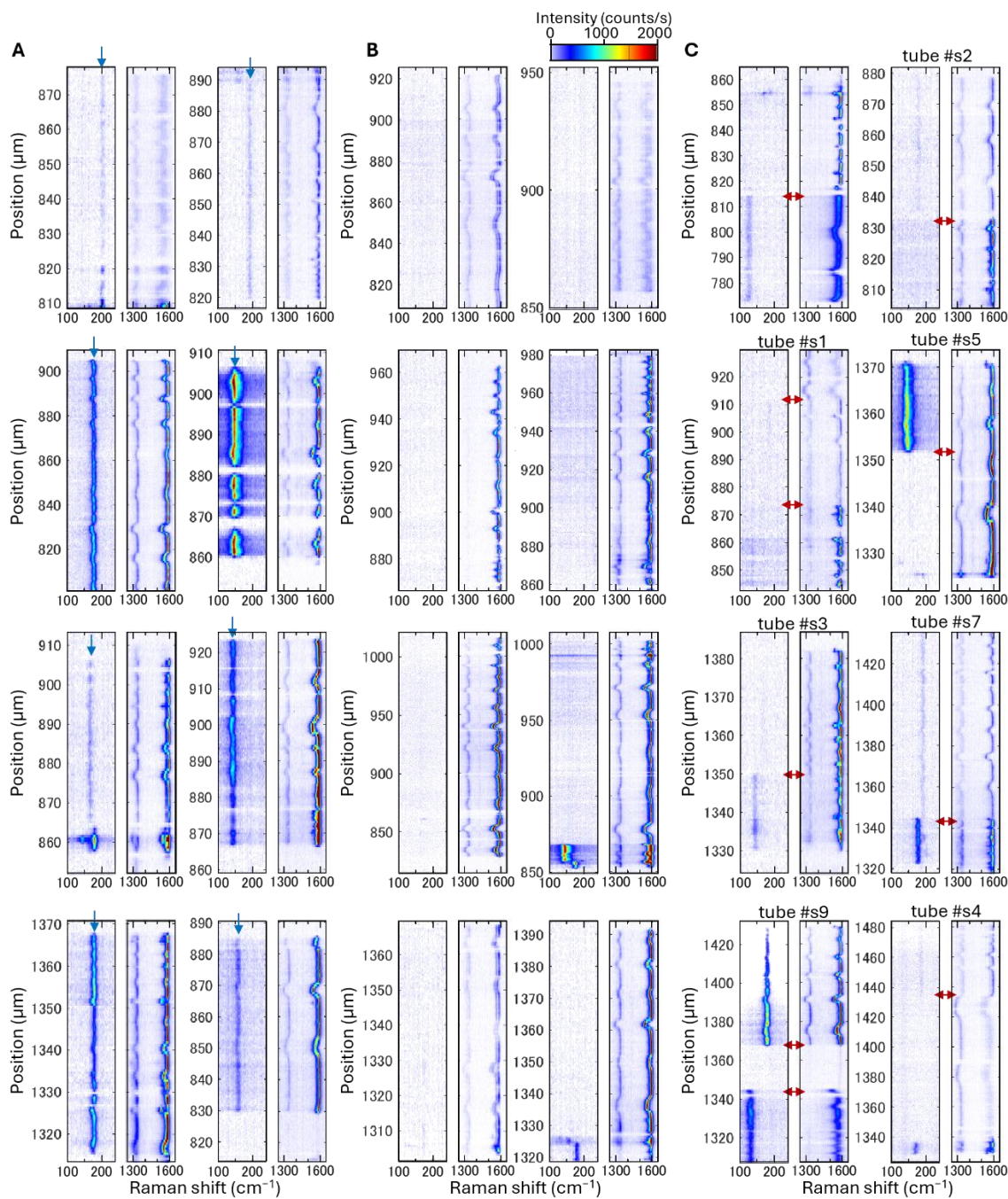

**Figure S5. Raman-mapping criteria for chirality preservation.** Raman spectra measured along individual CNTs at 600 nm intervals using an excitation wavelength of 532 nm. (**A**) CNTs exhibiting clear chirality preservation over their entire lengths, confirmed by the continuity of RBM features. (**B**) CNTs showing chirality preservation inferred with lower

confidence, based primarily on the continuity of G- and D-mode features. (**C**) CNTs exhibiting apparent intratube chirality changes, as evidenced by discontinuities in RBM frequencies, accompanied in some cases by changes in the G- and D-mode spectra. In all panels, Raman signals originating from the substrate (regions without CNTs) have been subtracted.

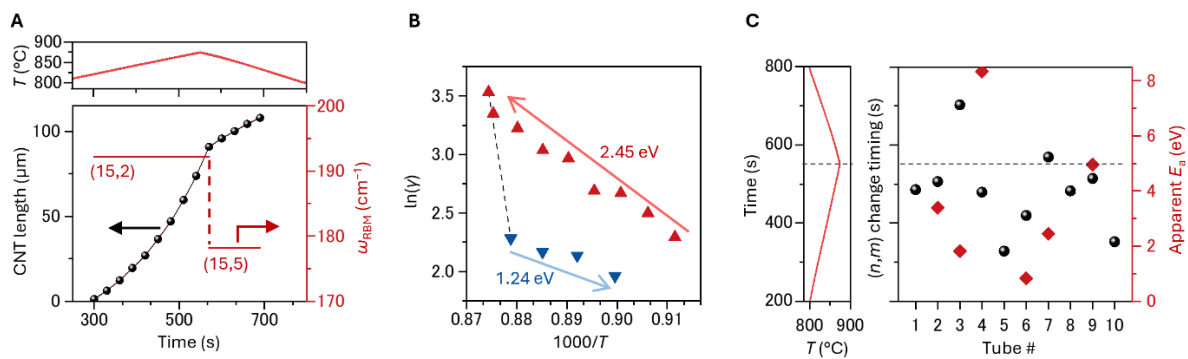

**Figure S6. Timing and kinetics of rare chirality-change events.** (**A**) Length evolution of the CNT (tube #7 in Fig. 5A) that shows a chirality change (bottom). Temperature profile is shown on top for comparison. (**B**) Arrhenius plots for this CNT, showing the different slopes and a large offset of growth rate $\gamma$. (**C**) Chirality-change timing and apparent activation energy before the transition for ten CNTs.

**Supplementary Note 7: Transfer and deposition of CNTs for Raman analysis**

CNTs grown on quartz substrates (Figure S7A) are not suitable for Raman analysis of isotope labels for several reasons. First, residual thermal stress induces non-uniform shifts in the Raman spectra (*5*). Second, the Raman scattering intensity is relatively weak on quartz. Third, the quartz substrate itself exhibits strong Raman peaks in the radial breathing mode (RBM) region. To mitigate these issues, the CNTs were transferred onto a $SiO_2$ (100 nm)/Si substrate pre-patterned with metal markers using a thin PMMA film (Figure S7B). A representative Raman map and the corresponding SEM image of the same area are shown in Figures S7C and S7D, respectively.

For FCCVD-derived CNTs, we performed similar Raman mapping measurements. Typically, CNTs synthesized via FCCVD are collected using membrane or glass-fiber filters; the ensemble Raman spectra shown in Figures 1 and S1 were obtained from such samples. However, as this study focuses on whether chirality remains continuous along a single CNT, it is essential to observe the nanotubes in an isolated morphology. Although isolation can be achieved by dispersing bundled samples from thick films into a solvent and then depositing them onto a substrate, this process involves a high possibility of cutting the CNTs, making them unsuitable for intended analysis along a single CNT. We therefore employed a direct dry deposition method via thermophoresis (*6*) during low-concentration CNT synthesis. Figure S8A shows an SEM image of the as-deposited CNTs. The G-mode intensity map for the region enclosed by the yellow rectangle is presented in Figure S8B. By tracing the Raman spectra along the axis of the CNT(s) indicated by the arrows, we observed approximately four distinct RBM peaks. Notably, the peak at 174 $cm^{-1}$ remains continuous from one end to the other, indicating that the chirality of this long CNT is preserved throughout its length. The other RBM peaks likely originate from short CNTs that are locally bundled or crossed with the long nanotube.

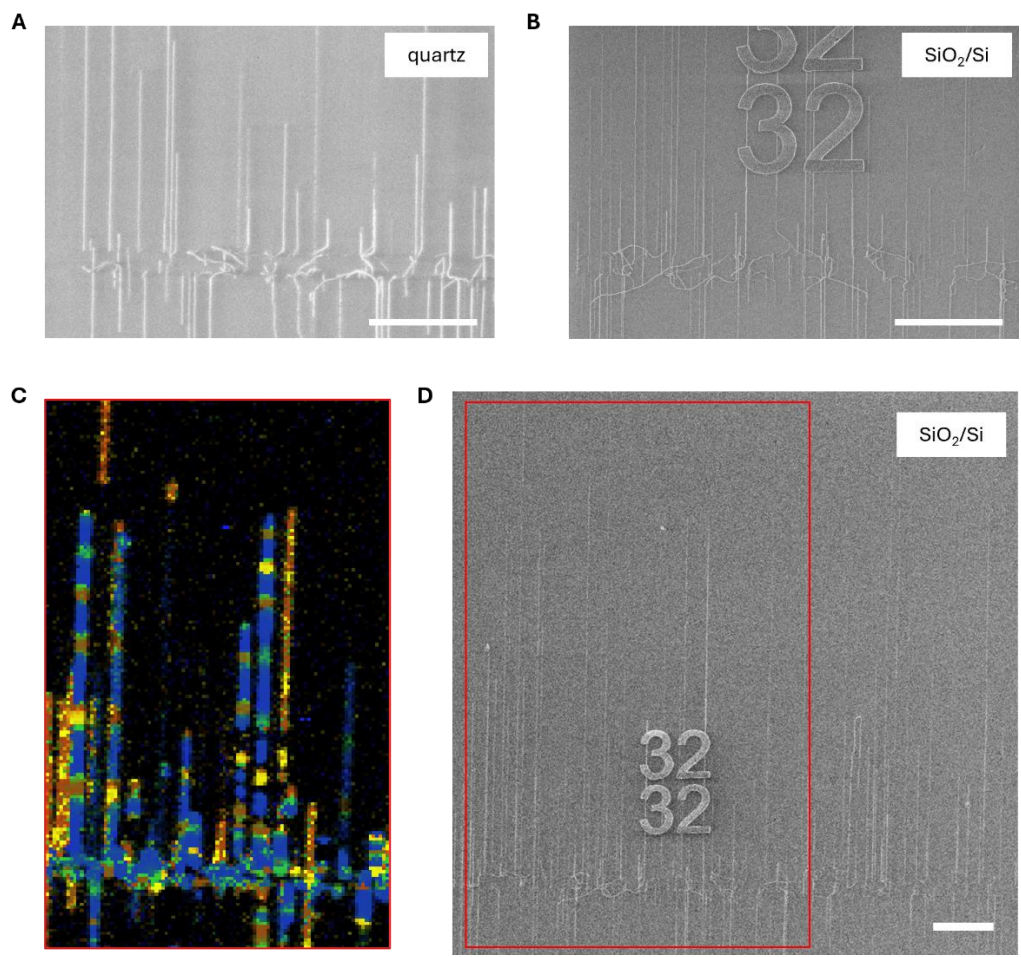

**Figure S7. Aligned, isolated CNTs for isotope labeling analysis.** (**A**) SEM image of as-synthesized carbon nanotubes (CNTs) obtained from an experiment with a temperature profile of 800-873-800°C. (**B**) SEM image of the CNTs on the same substrate after transfer onto a SiO$_2$/Si substrate for Raman analysis. (**C**, **D**) Corresponding Raman mapping image (C) and SEM image (D) of the identical area. In the Raman map, the color represents the G-mode peak frequency, and the brightness indicates its intensity. All scale bars are 10 μm.

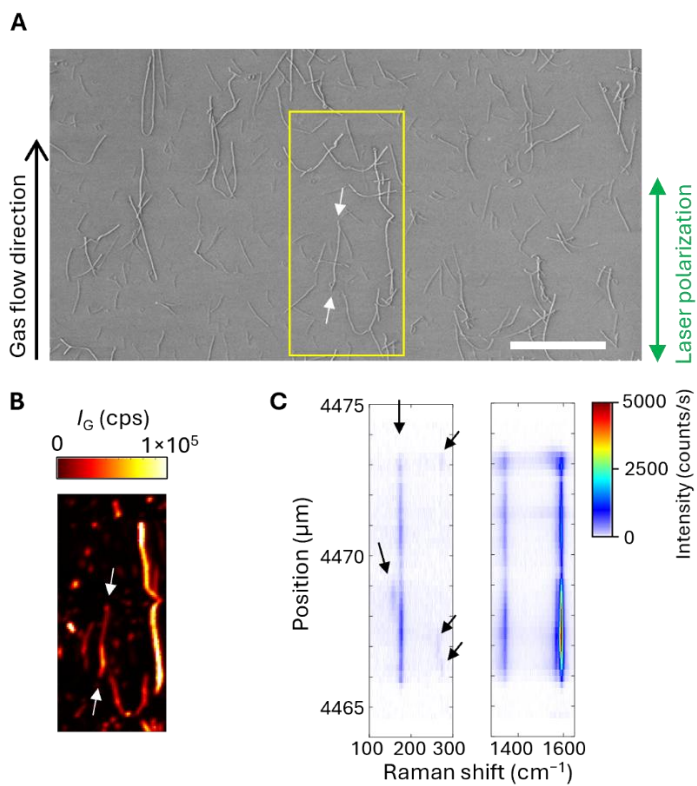

**Figure S8. As-grown CNTs on a substrate derived from FCCVD.** (**A**) SEM image of CNTs synthesized by FCCVD and directly deposited onto a SiO$_2$/Si substrate via thermophoresis. The gas flow direction is from the bottom to the top of the image. (**B**) Raman mapping image (integrated G-mode intensity) of the region enclosed by the yellow rectangle in panel A. The area corresponds to that shown in Fig. 5D of the main text. Excitation laser was linearly polarized parallel to the gas flow direction. (**C**) Raman spectra acquired along the axis of the CNT indicated by the arrows in panels A and B. The spectra reveal several short CNTs (approximately 1 μm) in close proximity to a longer CNT (>7 μm).